\documentclass{aa}
\usepackage[varg]{txfonts}
\usepackage{graphics}
\usepackage{natbib}
\bibpunct{(}{)}{;}{a}{}{,} 

\usepackage{hyperref}
\usepackage{url}
\renewcommand{\vec}[1]{\mathbf{#1}}

\hypersetup{colorlinks=true,citecolor=cyan,urlcolor=cyan,linkcolor=blue}

\begin{document}

\title{Tracing the ICME plasma with a MHD simulation}
\author{Ruggero Biondo\inst{\ref{UNIPA},\ref{OATO}}
\and Paolo Pagano\inst{\ref{UNIPA},\ref{OAPA}}
\and Fabio Reale\inst{\ref{UNIPA},\ref{OAPA}}
\and Alessandro Bemporad\inst{\ref{OATO}}}
\institute{University of Palermo, Physics and Chemistry Department, Piazza del Parlamento 1, 90134 Palermo, Italy\label{UNIPA}
\and INAF-Turin Astrophysical Observatory, via Osservatorio 20, 10025 Pino Torinese (TO), Italy\label{OATO}
\and INAF-Palermo Astronomical Observatory, Piazza del Parlamento 1, 90134 Palermo, Italy\label{OAPA}}

\abstract{The determination of the chemical composition of interplanetary coronal mass ejection (ICME) plasma is an open issue.
More specifically, it is not yet fully understood how remote sensing observations of the solar corona plasma during solar disturbances evolve into plasma properties measured in situ away from the Sun. 
The ambient conditions of the background interplanetary plasma are important for space weather because they influence the evolutions, arrival times, and geo-effectiveness of the disturbances.
The Reverse In situ and MHD APproach (RIMAP) is a technique to reconstruct the heliosphere on the ecliptic plane (including the magnetic Parker spiral) directly from in situ measurements acquired at 1 AU. It combines analytical and numerical approaches, preserving the small-scale longitudinal variability of the wind flow lines.
In this work, we use RIMAP to test the interaction of an ICME with the interplanetary medium. We model the propagation of a homogeneous non-magnetised (i.e. with no internal flux rope) cloud starting at 800 km/s at 0.1 AU out to 1.1 AU. Our 3D magnetohydrodynamics (MHD) simulation made with the PLUTO MHD code shows the formation of a compression front ahead of the ICME, continuously driven by the cloud expansion. 
Using a passive tracer, we find that the initial ICME material does not fragment behind the front during its propagation, and we quantify the mixing of the propagating plasma cloud with the ambient solar wind plasma, which can be detected at 1 AU.}


\maketitle


\section{Introduction}\label{sec:Introduction}

Since their discovery in the 1970s, coronal mass ejections (CMEs) and their interplanetary counterpart (ICME) have been a main topic for the heliophysics and the space weather community, being the most prominent transient structures in the heliosphere. ICMEs drive disturbances in the interplanetary medium and they are the only structures capable of producing extreme geomagnetic storms \citep{kilpua2017, luhmann2020}. Although often ambiguous \citep{richardsoncane2010}, the identification of ICMEs by 1 AU in situ measurements is possible thanks to certain signatures, which are generally detected in the parameters and distinguishable from the typical values of the ambient solar wind. These include stronger magnetic fields, enhanced plasma densities and speeds, lower proton temperatures, and suprathermal electron beams \citep{kilpua2017,richardsoncane2010,gosling1987,zwickl1983}. In addition, anomalies in the chemical composition and high charge states are often observed during ICME events. These signatures are of particular importance as they offer the possibility to directly observe material coming from the innermost layers of the solar atmosphere, inferring pre-eruption conditions \citep{rodriguez2016}.\\
One of the major open problems with ICMEs is understanding how the plasma properties observed in situ that are far from the Sun are related with those of the plasma originating in the inner corona (typically observed with remote sensing data) and/or to the subsequent interplanetary evolution and possible interaction with the interplanetary medium. This is also connected to the problem of understanding the chemical composition of ICME plasma and charge state distribution, a problem that must be addressed with numerical models alone \citep[e.g.][]{Lynch2011} or in combination with multi-spacecraft observations \citep[e.g.][]{Gruesbeck2011, reinard2012}. Moreover, tracing the chemical composition of solar phenomena is a very important issue in general, as the Sun is a very dynamic system, where the plasma is continuously remixed and its chemical composition can change locally with height \citep[from the photosphere to the corona, e.g.][]{Feldman1992}, geographically \citep[active versus quiet regions and solar wind, e.g.][]{Landi2015}, and temporally \citep[for instance during flares and eruptions, e.g.][]{Lepri2021, Brooks2021}. Spectroscopy usually provides important information about element abundances \citep[e.g.][]{Lee2012}, and it has been used, for instance, to track the presence of solar energetic particles (SEP) in ICMEs \citep{Brooks2021}, but in situ measurements would represent the best solution for directly measuring the composition of solar mass transported by ICMEs close to Earth. Studying the chemical composition of ICME plasma allows one to trace back its origin on the different layers of the solar atmosphere, also taking advantage of the so-called first ionisation potential (FIP) effect \citep[see recent review by][]{Song2020}. \\
In order to describe the possible mixing of the CME plasma accurately  with the ambient medium, it is crucial to reconstruct the ambient background condition of the heliospheric plasma since these conditions determine the interplanetary evolution of CMEs \citep[see][]{vanderholst2007, maloneygallagher2010, vrsnak2013, temmernitta2015} and the propagation of solar energetic particle streams \citep[e.g.][]{he2011}. Heliospheric magnetohydrodynamics (MHD) codes such as ENLIL \citep{odstrcil2009_ENLIL}, EUHFORIA \citep{pomoell2018}, and SUSANOO \citep{shiota2016_SUSANOO} are currently able to provide real-time forecasting of the space-weather conditions from the Sun to 1 AU and beyond. Nevertheless, the ability of these models to reconstruct the ambient conditions of the interplanetary plasma is still limited \citep[see e.g.][]{jian2016}, and this, for instance, has consequences on the reliability of solar disturbances forecasting \citep{riley2018}. Finer details have been shown to be reproduced by a recent approach \citep[][ more details in the following]{Biondoetal2021}.\\
In this work we use this approach to model the propagation of a CME reliably in the heliosphere. We also show that the CME plasma does not disperse and mostly maintains its composition as it crosses Earth's orbit, and we discuss to what extent we can expect the measured composition tracks to back the original one.

\section{The model}\label{sec:Description}\label{subsec:RIMAP_model}
The Reverse In-situ and MHD Approach (RIMAP) \citep[details in][]{Biondoetal2021} is aimed to bind MHD simulations of interplanetary medium to direct measurements of plasma parameters measured in situ at the distance of 1 AU. RIMAP combines, for the first time, the well known ballistic back-mapping approach \citep{Schatten1968, Wilcox1968} with a time-dependent MHD numerical simulation performed with the state-of-art PLUTO code \citep{mignone2007,mignone2012}. It provides a method to reconstruct the real-time conditions of interplanetary plasma from the Sun to 1 AU and beyond, preserving the smaller-scale density and velocity inhomogeneities dragged by the solar wind.
The computational grid used extends from $21.5 R_\odot$ ($\approx 0.1$ AU) up to $236.5 R_\odot$ ($\approx$ 1.1 AU) in the radial direction and covers the entire longitudinal interval, while the latitudinal aperture is restricted to $2^\circ$ centred around $\pi/2$. The time-dependent MHD equations are solved in a 3D spherical coordinate system corotating with the solar equator. The grid used has 256 cells in the radial direction (uniform with six cells up to 24.5 $R_\odot$ and then stretched with 250 cells up to the outer boundary), eight cells in the latitudinal direction, and 768 in the longitudinal one, both angular grids uniform.
We used the conservative approach to maximise the conditions of CME plasma mixing with the interplanetary medium in the ecliptic plane. Accordingly, we chose to describe an ejected cloud with no internal magnetic flux rope that is embed in the ambient magnetic field  \citep[non-MC ICME,][]{kilpua2017} and to assume that in the early phase the angular width and the velocity of the perturbation remain constant. The initial ICME can then be described by an ice-cream cone model \citep[e.g.]{zhao2002,xie2004,xue2005,gopalswamy2009,hyeonock2017}: a homogeneous plasma cloud, isotropic in expansion and with radial bulk velocity. The small number of input parameters needed by CME cone models make them particularly convenient for routine applications in space weather forecasting\citep{scolini2018} despite the drawback of not including an internal magnetic field.\\
The kinematic parameters of a cone-modelled ICME are determined by its angular width $\omega$ and radial speed $v_{ICME}$ \citep{xie2004}. Assuming a nearly circular cross section, the ICME can be thought of (before entering the computational domain) as a spheroid pushed at a constant speed, and it enters at the 0.1 AU inner boundary as a time-dependant boundary condition \citep[e.g.][]{odstrcil_pizzo1999a,pomoell2018}. The angle subtended by the spheroid with the inner boundary surface during its passage is $\alpha(t) = \frac{\omega}{2}\sin\left(\frac{\pi}{2}(t-t_0)\frac{v_{ICME}}{R_b}\sin^{-1}\frac{\omega}{2}\right),$ where $R_b$ is the radius of the inner boundary and $t_0$ is the onset time of insertion into the domain. Furthermore, $\alpha(t)$ is used to determine which cells on the longitudinal axis of the inner radial boundary belong to the ICME: if $(\varphi-\varphi_0)^2\leq\alpha(t)^2$ holds true, the background solar wind parameters are replaced with the perturbation ones. The model assumes that the ICME plasma cloud propagates homogeneously before entering the simulation domain. The velocity $\vec{v}_{ICME}$ at each point is therefore parallel to the purely radial velocity of the cells found at the longitude corresponding to the insertion $\varphi_0$, and therefore it has latitudinal and longitudinal components that are non-zero and dependent on $\theta$ and $\varphi$. The number density and plasma temperature inside the ICME are also homogeneous, with values chosen close to the average event ones during solar cycle 23 and 24 \citep[][see table \ref{table:CME_parameters}]{gopalswamy2014}.

Finally, an artificial, passive scalar is added to the ICME model and used as a tracer of the CME plasma flow because it is simply advected and interacts in no way with the physical quantities of the model. Before the insertion time $t_0$, this tracer is null throughout the computational domain. After $t_0$, all the cells satisfying $\Delta\varphi^2\leq\alpha(t)^2$, that is within the CME cone, are filled with 1.
During the evolution, the tracer can flow in and out of each grid cell of the simulation domain, thus attaining a mixing with the background plasma at the grid length scale.

\section{Model results}\label{sec:Results}
\label{sec:ICME_results}
\begin{figure}[ht]
\begin{center}
\includegraphics[trim=0 0 0 0, clip,width=0.5\textwidth]{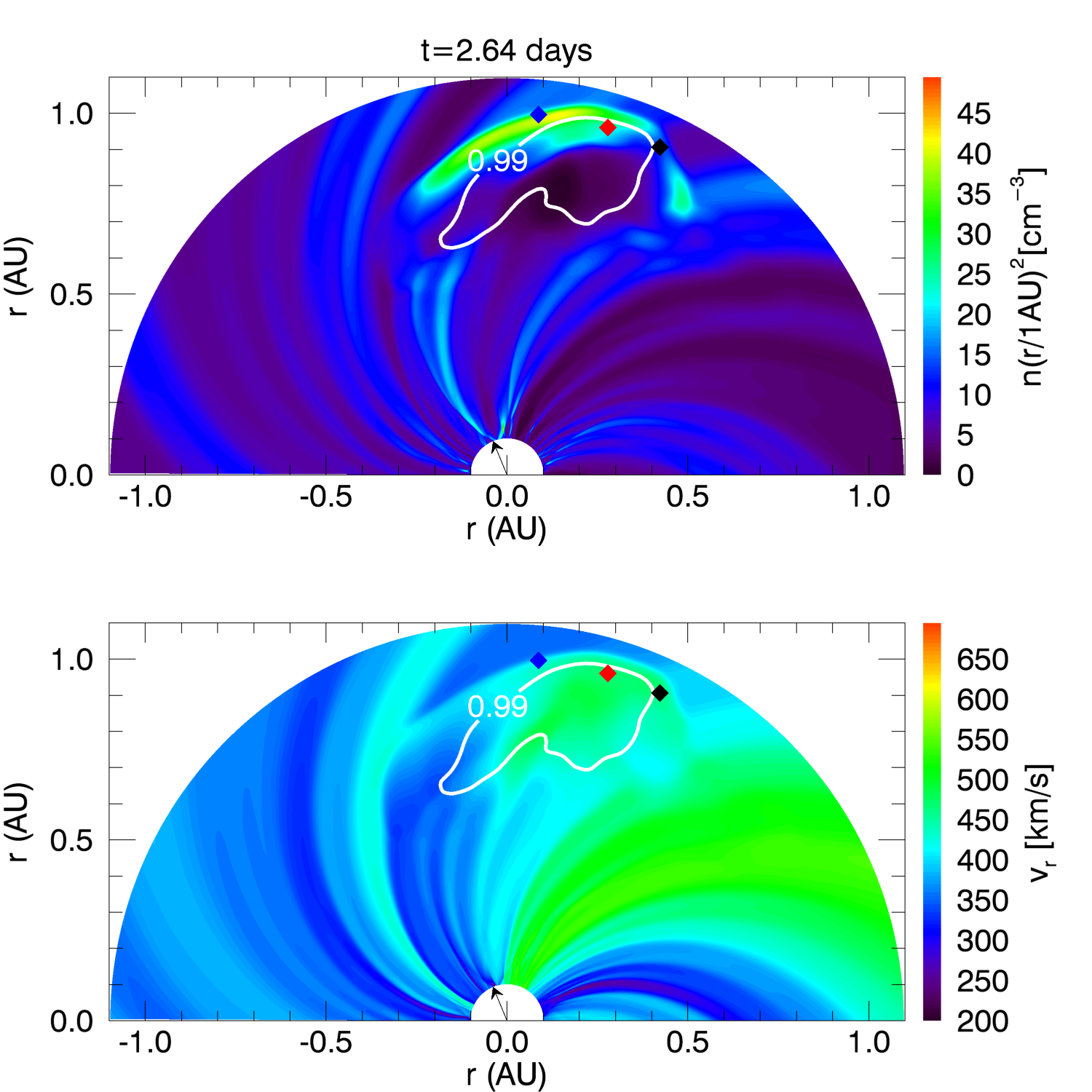}
\caption{Contour (99\%) of the ICME tracer mass (white) on colour maps of plasma number density and radial velocity at the labelled time of the MHD simulation (see text for details). The coloured diamonds ($\varphi=85^\circ,\ 73.8^\circ$,  $65^\circ$) show the position of three virtual spacecrafts, which would detect the time profiles in Fig. \ref{fig:time_profiles}. The black arrow indicates the entry point of the ICME into the inner boundary. See attached animation.}
\label{fig:ICME_n_tr_230}
\end{center}
\end{figure}

The background condition of the interplanetary Parker spiral for the ICME simulation is reconstructed by RIMAP using plasma parameters measured by Wind in 2009 between March 3 and March 29 \citep[][]{Biondoetal2021}, around the minimum of solar activity cycle 23. Even if this time frame was chosen due to its relatively calm solar wind conditions, the RIMAP-reconstructed Parker spiral is highly structured in its longitudinal variability (Fig.\ref{fig:ICME_n_tr_230}), offering diverse conditions to the ICME propagation. At $t_0=0$, the cloud enters the inner radial boundary at a longitude of $\varphi_0=112.5^\circ$ at the location of a relatively high-density streamline (arrow in Fig.\ref{fig:ICME_n_tr_230}). This means that we are addressing the impact of an ICME on a dense, slow stream that may result in the interplanetary formation of geo-effective disturbances.  The parameters are in table \ref{table:CME_parameters}.
\begin{table}
\caption{ICME input parameters at 0.1 AU} 
\begin{tabular}{l l}
\hline
Angular width $\omega$ &  15$^\circ$\\
Radius $r_{ICME}$ &  5.76 R$_\odot$\\
Bulk speed $v_{ICME}$ &  800 km/s\\
Number density $n_{ICME}$ &  600 cm$^{-3}$\\
Temperature $T_{ICME}$ & $8\cdot10^5$ K\\
\hline
\label{table:CME_parameters}
\end{tabular}
\end{table}
\\
Fig.\ref{fig:ICME_n_tr_230} shows ecliptic maps of plasma density and radial speed in a frame co-rotating with the solar equator at a time in which the ICME has propagated from the insertion point at 0.1 AU almost to the outer domain boundary (1.1 AU). Even if the cloud is homogeneous in density and speed before entering the inner radial boundary, it is soon stretched and slowed down by the interaction with the interplanetary medium. The speed drops to less than 500 km/s at 1 AU and the CME trajectory has been deflected from being the purely radial. This deceleration to the ambient speed of solar wind is a common feature in observed ICME events \citep{kilpua2017, gopalswamy2000, lindsay1999} and simulated ones \citep[e.g.][]{manchester2014}.
In the early stages of the propagation, a compression front forms ahead of the cloud and  develops an internal structure due to the inhomogeneity of the background solar wind streams. In particular, from day 1 onward, the right edge of this front begins to detach from the rest and at t=2.5 days it forms a separate blob.
At the same time, behind the front, the cloud is expanding, its density decreases, and a void forms, where density is one-tenth of the ambient one at 1 AU.

In Fig.\ref{fig:ICME_n_tr_230} the contours of the passive tracer $T_r$ measure the mixing of the original perturbation plasma with the perturbed ambient plasma. Its mass percentage is calculated as
\begin{equation}\label{eq:tracer_percentage}
T_{rg}(r,\varphi) = \frac{\int d\theta\ r\ T_r(r,\theta,\varphi)\  \rho(r,\theta,\varphi)}{\int d\theta\ r\ \rho(r,\theta,\varphi)} \times 100
.\end{equation}
In Fig.\ref{fig:ICME_n_tr_230}  each white contour bounds the area where $99\%$ of the tracer mass is contained and interestingly shows that, in spite of the interaction with the different arms of the Parker spiral, the original material of the ICME remains compact behind the front as a single coherent structure, and it is more co-spatial with the velocity CME field (green 'island' in the bottom panel). This result may have important implications regarding the chemical composition of ICME plasma measured with in situ data.\\
In Fig.\ref{fig:ICME_n_tr_230} we mark three locations at 1 AU, which are along the ICME path and also in solar wind streams with different properties. In order to study how the ICME passage affects in situ measurements, we analyse the evolution of the plasma parameters and magnetic field at these locations as a function of time. This mimics the configuration in which multiple spacecrafts detect the interplanetary plasma at different locations along Earth's orbit and/or along the CME passage. The red diamond (hereafter, position A) is taken where the tracer has its maximum on the domain, that is where there is the largest percentage of original ICME plasma at the time when the perturbation has moved to 1 AU. The blue diamond (position B) is an intermediate distance between the ICME centre and its boundaries (at 1 AU). The black diamond (position C) falls almost outside of the outer ICME contour line of the tracer and it is hit only marginally by the perturbation front.

\begin{figure}[ht]
\centering
\includegraphics[trim=0 0 0 0, clip,width=0.5\textwidth]{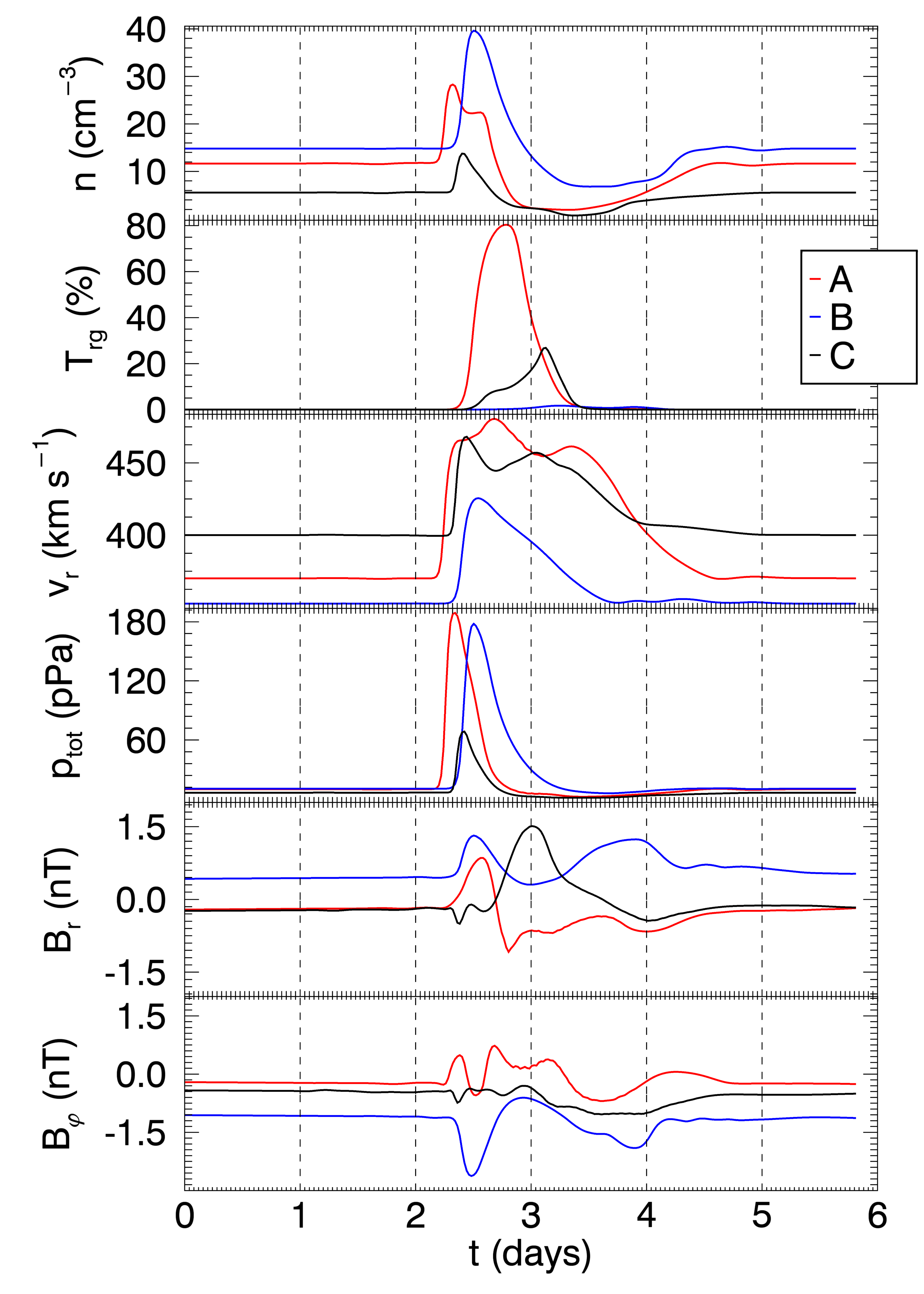}
\caption{Time profiles of number density, percentage of tracer, radial speed, pressure,  and radial and longitudinal component of the magnetic field at three different 1-AU positions, A (red), B (blue), and C (black), marked in Fig.\ref{fig:ICME_n_tr_230}. Here the ICME enter time is $t_0=0$.}
\label{fig:time_profiles}
\end{figure}

Fig.\ref{fig:time_profiles} shows the time profiles (from top to bottom) of the plasma density, tracer mass percentage, radial speed, total pressure, and the radial and longitudinal magnetic field,  taken from the MHD simulation that would be measured at position A (red), B (blue), and C (black) lines. All quantities are averaged on the latitudinal domain.\\
Approximately 2.5 days after the insertion time, a rapid increase in the density, pressure, speed, and magnetic field is detected at 1 AU in A, B, and C, marking the beginning of the ICME event. Plasma density reaches a higher peak value ($\sim40$ cm$^{-3}$) at position B, which is somewhat later than at positions A and C. This happens because, as can be seen in the animation attached to Fig.\ref{fig:ICME_n_tr_230}, the perturbation front which hits position B is more compressed against a high density stream, while at position A the front has travelled through more tenuous streams and thus accumulated less plasma. Position C is just outside of the cloud boundaries and on the trajectory of the region where the front has split into two, and there we see only a small and short-lasting bump of a few cm$^{-3}$, all before $t=3$ days, followed by a longer-lasting slight decrease back to the unperturbed ambient value at $t=5$ days. The density bump lasts much less than 1 day at all positions. The total pressure shows similar and simultaneous bumps. After the bump, we still see a density perturbation in the form of a broad dip, which brings the density down to about 0 at position B and C, and it lasts about 1 day or more.\\
Interestingly, the tracer evolution shows a significantly delayed bump at positions A and even more at C, and almost no change at position B. The tracer peak at position A tells us that there, about 80\% of the plasma consists of the original ICME plasma. On the other hand, this peak takes place later than the peak of total density, when the density is already steeply decreasing, that is, in the tail of the perturbation. At position C, the fraction of original ICME plasma has a much lower peak ($\sim 30$\%), and this occurs even later, close to the density dip. Overall, while position A is close to the ICME core and rich in original ICME mass, and while C is intermediate, position B is still significantly perturbed, yet devoid of a tracer. Contrary to common sense, we clearly see that the original CME plasma is well detached from the compression front moving ahead of it, which is visible in the density and pressure plots, and it instead permeates the void behind it.\\
The radial speed shows a much longer-lasting bump, quite similar at the three positions. In both A and C, the plasma speed jumps by $\sim 150-200$ km/s from their pre-event values up to to $\sim450-500$ km/s, and it remains at these speeds for most of the two days of the perturbation. As Fig.\ref{fig:ICME_n_tr_230} shows, the perturbation is quite homogeneous in velocity even at 1 AU and despite being just outside of the bulk of the ICME, C still shows a velocity jump similar to A and B. On the other hand, at position B, the perturbation hits higher density wind streams and decelerates more rapidly to the pre-event value in about 1 day. \\
As mentioned before, the total pressure shows sharp peaks similar to the density ones, up to about 180 pPa, lasting less than one day, while in C the bump is much smaller, up to only about 60 pPa. This is consistent with the fact that A and B are directly hit by the full perturbation front, while in C the front is attenuated by the interaction with the denser wind stream.\\
Despite the starting cloud having no internal magnetic flux rope, the simulated ICME perturbs the magnetic configuration at 1 AU, as shown by the $B_r$ and $B_\varphi$ profiles. Both components show significant fluctuations, significantly out of phase and significantly different at the three positions, which is evidence of a strong distortion and compression of the field lines. 

\section{Conclusions}\label{sec:Conclusions}

In this work, with the RIMAP-MHD model \citep{Biondoetal2021}, we simulate the propagation of an ICME inside a background solar wind reconstructed reliably from the measurements acquired in situ around the minimum of solar activity cycle 23. The original ICME plasma is tracked by using a passive tracer, and its arrival at 1 AU is sampled in situ by virtual spacecrafts located at three different longitudes. We show that, although the complex interaction with multiple fast and slow wind streams leads to ICME non-uniform deceleration and deformation, the original ICME bubble remains compact and is not eroded, thus preserving its chemical composition. On the other hand, the perturbation front is almost entirely composed of interplanetary plasma compressed during its propagation against the pre-ICME background. These results have a potential impact on the interpretation of multi-spacecraft observations of solar eruptive phenomena.

This work sheds some light on how CMEs observed near the Sun evolve, becoming ICMEs that are detected in interplanetary space. In the past, after some first attempts by using test particles \citep[][]{Riley1997, Odstrcil1999}, a real particle tracking method was applied to a MHD simulation of an ICME, for instance, by \citet{Riley2008}. In that work, it was shown that in the interplanetary evolution, the counterpart of the CME front is buried within the sheath material of the ICME, which formed by upstream solar wind plasma swept up by the CME.
However, using a tracer for which we describe the advection in the MHD code instead of sampling particles, we can provide a quantitative analysis of this effect.
We find that only a very small fraction of the original CME plasma is present in the compression front, but also that this fraction can rise up to about 80\% at other locations and times during the event. Our modelling thus confirms  that strong variations of chemical composition measured at 1 AU can directly trace back, at least in part, to the original CME composition.

More recently, \citet{Brooks2021} have identified the source region of the energetic particles during an eruption. Again, our model corroborates and explains those findings as it shows that the ICME can maintain a distinguished chemical composition during its motion in the interplanetary space with a limited mixing of the plasma.

\begin{acknowledgements}
This work acknowledges support from ASI/INAF Solar Orbiter contract n. 2018-30-HH.0 and from the Italian Ministero dell’Università\'{a} e della Ricerca (MUR).
\end{acknowledgements}

\bibliographystyle{aa} 
\bibliography{rimap2} 

\end{document}